\documentclass[sn-mathphys,iicol,pdflatex]{sn-jnl}

\usepackage{graphicx}%
\usepackage{multirow}%
\usepackage{amsthm}%
\usepackage{mathrsfs}%
\usepackage[title]{appendix}%
\usepackage{textcomp}%
\usepackage{manyfoot}%
\usepackage{booktabs}%
\usepackage{algorithm2e}
\usepackage{listings}%
\usepackage[T1]{fontenc}             
\usepackage[latin1,utf8]{inputenc}  
\usepackage{amsmath,amssymb,amsfonts,amsbsy,bbm}
\usepackage{pstool}
\usepackage{caption}
\usepackage{setspace}
\usepackage{bm}
\usepackage{upgreek}
\usepackage{float}
\usepackage{subfig}
\usepackage{multirow, bigstrut}
\usepackage{color,soul}
\usepackage[table]{xcolor}
\definecolor{tabcolor}{rgb}{0.9,0.95,0.9}
\definecolor{headcolor}{rgb}{0.8,0.9,0.8}
\newcolumntype{a}{>{\columncolor{tabcolor}}c}
\newcolumntype{d}{>{\columncolor{headcolor}}c}

\newcommand{\tensor}[1]{\mbox{\boldmath ${#1}$}}

\DeclareCaptionType{mycapequ}[][List of equations]
\DeclareMathAlphabet\mathbfcal{OMS}{cmsy}{b}{n}

\renewcommand{\[}{\left[}
\renewcommand{\]}{\right]}
\renewcommand{\(}{\left(}
\renewcommand{\)}{\right)}

\newcommand{\grad}{\nabla}

\newcommand{\of}[1]{\left( {#1} \right)}

\newcommand{\IntBo}{\int_{\cB_0}\ }
\newcommand{\IntBoh}{\int_{\cB_0^h}\ }
\newcommand{\IntBoe}{\int_{\cB_0^e}\ }
\newcommand{\IntBt}{\int_{\cB_t}\ }

\newcommand{\IntSBo}{\int_{\partial\cB_0}\ }

\newcommand{\IntSBoh}{\int_{\partial\cB_0^h}\ }

\newcommand{\dV}{\text{\:dV}}
\newcommand{\dv}{\text{\:dv}}
\newcommand{\dA}{\text{\:dA}}

\DeclareMathOperator*{\asse}{\scalerel*{\text{A}}{\textstyle\sum}}
\usepackage{scalerel}

\newcommand{\cB}{\mathcal{B}}
\newcommand{\cF}{\mathcal{F}}

\newcommand{\cN}{\mathcal{N}}
\newcommand{\cP}{\mathcal{P}}
\newcommand{\cR}{\mathcal{R}}
\newcommand{\cG}{\mathcal{G}}

\newcommand{\cJ}{\mathcal{J}}

\theoremstyle{thmstyleone}%
\theoremstyle{thmstyletwo}%
\newtheorem{remark}{Remark}%

\theoremstyle{thmstylethree}%

\begin{document}

\title[Article Title]{Configurational-force-driven adaptive refinement and coarsening in topology optimization}


\author*[1]{\fnm{Gabriel} \sur{Stankiewicz}}\email{gabriel.stankiewicz@fau.de}
\author[1]{\fnm{Chaitanya} \sur{Dev}}\email{chaitanya.dev@fau.de}
\author[1]{\fnm{Paul} \sur{Steinmann}}\email{paul.steinmann@fau.de}

\affil*[1]{\orgdiv{Institute of Applied Mechanics}, \orgname{Friedrich-Alexander-Universit\"at Erlangen-N\"urnberg}, \orgaddress{\street{Egerlandstr. 5}, \city{Erlangen}, \postcode{91058}, \state{Bavaria}, \country{Germany}}}


\abstract{The iterative nature of topology optimization, especially in combination with nonlinear state problems, often requires the solution of thousands of linear equation systems. Furthermore, due to the pixelated design representation, the use of a fine mesh is essential to obtain geometrically well-defined structures and to accurately compute response quantities such as the von Mises stress. Therefore, the computational cost of solving a fine-mesh topology optimization problem quickly adds up. To address this challenge, we consider a multi-level adaptive refinement and coarsening strategy based on configurational forces. Configurational forces based on the Eshelby stress predict configurational changes such as crack propagation or dislocation motion. Due to a relaxation in the calculation of (Eshelby) stresses with respect to the design variables, discrete configurational forces increase not only in highly stressed regions, but also in grey transition regions (design boundaries). For this reason they are an ideal criterion for mesh adaptivity in topology optimization, especially when avoiding stress failure is a priority. By using configurational forces for refinement, we obtain a high-resolution structure where the refined mesh is present along the design boundaries as well as in stress-critical regions. At the same time, multilevel coarsening using the same criterion drastically minimizes the computational effort.}

\keywords{topology optimization, adaptive refinement, adaptive coarsening, configurational forces}



\maketitle

\section{Introduction}

The rapid development of computational resources has enabled the execution of increasingly complex and numerically demanding simulations, even on personal computers. These advances in hardware have been accompanied by intensified research on computational methods, which has led to an increasing role for computer simulations in product development. This is due to the low costs, reliability, accuracy, and efficiency of such simulations. However, the fundamental challenge remains the trade-off between solution accuracy and computational efficiency, which is exemplified by the degree of discretization of the system to be solved. It is evident that the finer the discretization, the more accurate the solution. However, this increased accuracy is associated with a higher computational cost, regardless of the specific computational method employed. This is particularly evident in topology (or any structural) optimization, where linear systems need to be solved hundreds of times, sometimes in a nested fashion, e.g., for each optimization iteration multiple load steps and Newton-Rhapson iterations are required for highly nonlinear boundary value problems.

Therefore, a relevant research direction in the world of computer simulation is focused on discretization techniques that aim to maximize the accuracy/cost ratio. In the context of the finite element method (FEM), many commercial programs offer advanced preprocessing tools with different meshing algorithms.  However, these almost exclusively produce unstructured meshes, which introduce an additional level of complexity in topology optimization as compared to the original approach based on a structured mesh. For simulations involving iterative solving of linear systems, i.e. quasi-static nonlinear elastic problems or time discretized problems, there are several techniques that modify the mesh for each iteration.

In the context of topology optimization, mesh adaptivity is additionally focused on maximizing geometric accuracy, furthermore to the ratio of numerical accuracy and computational cost. Several works deal with mesh refinement strategies for tria-based meshes \citep{costa2003layout, stainko2006adaptive, nana2016towards}, (classical) structured meshes with hanging node constraints (with coarsening) \citep{wang2010dynamic, salazar2018adaptive, salazar2020three, munoz2022improvement} or unstructured quad-based meshes \citep{lambe2018topology}. The adaptivity criteria can be primarily divided into density-based \citep{wang2010dynamic, nana2016towards, lambe2018topology}, which prioritize the geometric accuracy of the designs, and error-based \citep{salazar2018adaptive, zhang2020adaptive}, which prioritize the computational accuracy. The work of \cite{munoz2022improvement} tackles both density-based and error-based adaptivity to address both geometry representation and accurate stress computation to meet the constraints. In combination with shape optimization employing an embedding domain discretization (EDD), adaptive meshing has been applied to the variable shape in \cite{stankiewicz2021coupled}. Moreover, in \cite{dev2022sequential, stankiewicz2022geometrically, stankiewicz2024towards} mesh adaptivity in the EDD-based shape optimization was introduced for both the structured mesh of the computational domain and the variable shape.

In the following, we employ the concept of configurational forces for mesh adaptivity in topology optimization. Configurational (or material) forces are a well-established driving criterion for fracture propagation \citep{maugin1995material, steinmann2000application, miehe2007robust, kuhn2016discussion, ballarini2016newtonian, moreno2024configurational}. Furthermore, configurational forces have been successfully explored for r- and h-adaptation of meshes in the context of hyperelasticity \citep{mueller2004use}, poromechanics \citep{na2019configurational}, phase field fracture \citep{welschinger2010configurational}, isogeometric analysis \citep{henap2017configurational}, elastoplasticity \citep{henap2014numerical}, fracture mechanics \cite{qinami2019circumventing} or truss optimization \citep{askes2005structural}. In this article, we show that configurational forces are also a powerful criterion for mesh adaptivity in topology optimization and offer a promising alternative to density-based and stress-based adaptivity techniques. We propose a robust adaptivity strategy that utilizes the deal.II finite element library to handle refinement and coarsening with hanging node constraints \citep{bangerth2007deal}. 

This article is organized as follows. In section \ref{sec:config} we motivate the use of configurational mechanics and re-iterate its essential concepts. We also discuss the meaning and role of configurational forces. In section \ref{sec:method} we introduce our methodology for mesh adaptivity using configurational forces and adapt the configurational forces to incorporate the pseudo-density variables. We address the possible edge cases and how deal.II handles mesh coarsening and refinement. In section \ref{sec:examples}, two examples, a cantilever beam and a U-beam, are studied to evaluate the introduced mesh adaptivity based on configurational forces. A thorough comparison with other known adaptivity strategies is given. Concluding remarks are given in section \ref{sec:conclusions}.

\section{Why configurational mechanics?}\label{sec:config}

In the conventional understanding of continuum mechanics, considering finite deformations, we are concerned with finding the spatial deformation of a continuum under spatial (deformational) forces, be it boundary tractions or a body-distributed force, e.g. gravity. By assuming quasi-static loading conditions, we distinguish between the material (reference) configuration of such a system, which corresponds to the underformed (unloaded) state of a continuum body, and the spatial configuration that is in equilibrium for the external loads. For an elastic structure this equilibrium corresponds to a minimum of the total potential energy, i.e. the sum of the internal energy and external potential energies.

The core assumption of conventional deformational continuum mechanics is that the material configuration is fixed, i.e. it does not allow any geometrical changes. In contrast, configurational continuum mechanics not only allows for such configurational changes, but in fact is primarily concerned with them. In other words, the material configuration is no longer fixed and allows energy release (in accordance with the second law of thermodynamics) in the form of configurational changes such as crack propagation or defect motion. 

Analogously, if spatial (deformational) forces are the driving mechanism behind the spatial deformation of a body, we recognize material (configurational) forces that are the driving mechanism behind any configurational changes. Configurational forces provide us with the necessary information to understand what are the energetic implications resulting from configurational changes such as crack propagation or dislocation motion. This information is highly useful in determining regions where the risk of such configurational change is highest. However, the application of configurational forces is more versatile than merely the prediction of material defects.  Therefore, in this article, we demonstrate the power of configurational forces as a criterion for finite element mesh adaptivity in the context of topology optimization. 

\subsection{Essential concepts}

To provide a thorough understanding of configurational forces, basic concepts from classical deformational continuum mechanics for the spatial (deformational) problem are first reviewed. Subsequently, the analogous quantities for the material (configurational) problem are introduced, which form the core of configurational mechanics.

\subsubsection{Deformational problem}

\begin{figure}[H]
	\centering
	\includegraphics[width=84mm]{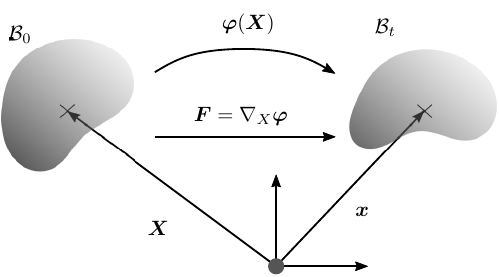}{}
	\caption{Basic kinematics of the deformational problem in continuum mechanics.}
	\label{fig:spatialmotion}
\end{figure}

Let $\cB_0$ (Fig. \ref{fig:spatialmotion}) be material configuration of a continuum body (subscript $0$ indicates the material configuration). The body $\cB$ is composed by the continuum points $\cP$ with $\cB = \{\cP\}$. Then the material configuration results from the material placement $\tensor{X}$ of every $\cP$ into $\mathbb{E}^3$. In response to spatial forces, in the form of tractions $\tensor{t}_0$ and body forces $\tensor{b}_0$, the body undergoes deformation described by the spatial deformation map $\tensor{\varphi} (\tensor{X})$ resulting in the spatial configuration $\cB_t$. The body in the spatial configuration, which is the actual deformed state of the body, is described by the spatial placement $\tensor{x} = \tensor{\tensor{\varphi}} (\tensor{X})$.

In classical deformational continuum mechanics of solids, all spatial quantities, their gradients and variations are usually evaluated with respect to the fixed material placement $\tensor{X}$\footnote{Alternative formulation, where all spatial quantities, their gradients and variations are evaluated with respect to the spatial placement $\tensor{x}$, is typical for fluid dynamics, where due to the nature of fluids no material reference body exists.}. Thus, the spatial deformation gradient is defined as $\tensor{F} = \grad_X \tensor{\varphi}$.

To obtain the balance equations for the deformational problem, we introduce the total energy density $U_0(\tensor{\varphi},\tensor{F};\tensor{X})$ consisting of the internal potential $W_0(\tensor{F};\tensor{X})$ per unit reference volume and the external potential $V_0(\tensor{\varphi};\tensor{X})$ per unit reference volume. The total potential energy is given by
\begin{equation}
\begin{split}
\it\Pi(\tensor{\varphi};\tensor{X}) 
&= \IntBo U_0(\tensor{\varphi},\tensor{F};\tensor{X}) \dV \\
&= \IntBo W_0(\tensor{F};\tensor{X}) + V_0(\tensor{\varphi};\tensor{X}) \dV.
\end{split}
\end{equation}

The common virtual work statement is then obtained via the stationarity condition, which requires the spatial variation of the total potential energy at fixed material placement to vanish, i.e. $\delta_X {\it\Pi} = 0$, for any admissible test function $\delta \tensor{\varphi}$
\begin{equation}
\begin{split}
\delta_X \it\Pi = \IntBo \partial_F U_0 : \grad_X \delta \tensor{\varphi} \dV + \IntBo \partial_\varphi U_0 \cdot \delta \tensor{\varphi}  \dV \\
= \IntBo \tensor{P} : \grad_X \delta \tensor{\varphi} \dV - \IntBo \tensor{b}_0 \cdot \delta \tensor{\varphi} \dV = 0,
\end{split}
\end{equation}
where the derivative $\partial_F U_0 =: \tensor{P}$ yields the two-point Piola stress tensor and the derivative $\partial_\varphi U_0 =: -\tensor{b}_0$ is the spatial body force. In localized form, using Cauchy's theorem, $\tensor{t}_0 := \tensor{P} \cdot \tensor{N} \equiv \tensor{0}$\footnote{Here we assume zero traction BCs for the sake of presentation. A more complete account on configurational mechanics including boundary potentials may be found in \cite{steinmann2022spatial}.}, where $\tensor{N}$ is the outward unit normal to the reference boundary $\partial\cB_0$, we obtain the localized equilibrium of deformational forces
\begin{equation}
\text{Div} \tensor{P} + \tensor{b}_0 = \tensor{0},
\label{eq:localbalancep}
\end{equation} 
where the divergence operator $\text{Div} \{\bullet\}$ is evaluated with respect to the material placement $\tensor{X}$.

\subsubsection{Configurational problem}

\begin{figure}[H]
	\centering
	\includegraphics[width=84mm]{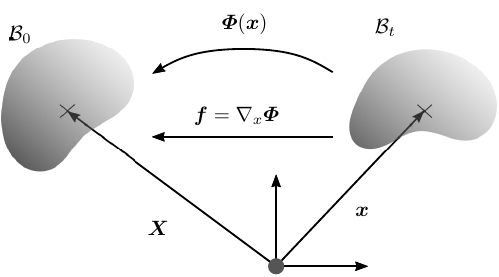}{}
	\caption{Basic kinematics of the configurational problem in continuum mechanics.}
	\label{fig:materialmotion}
\end{figure}

Now, to reiterate, the essence of configurational mechanics lies in the consideration of configurational changes. Here we consider the spatial placement of the body $\tensor{x}$ (Fig. \ref{fig:materialmotion}) as fixed and search for the change of the material configuration, described by the material deformation map $\tensor{X} = \tensor{\varPhi}(\tensor{x})$, analogously parametrized in the spatial placement $\tensor{x}$. Naturally, we consider the material deformation gradient $\tensor{f} = \grad_x \tensor{\varPhi} = \tensor{F}^{-1}$, where the gradient operator $\grad_x \{\bullet\}$ is evaluated with respect to the fixed spatial placement $\tensor{x}$. 

Next, we re-parametrize and re-express the total potential energy density $U_t(\tensor{\varPhi},\tensor{f}; \tensor{x})$ per unit volume in the spatial configuration so that $U_0 = \det (\tensor{F}) U_t$. The total potential energy is then given by 

\begin{equation}
\pi(\tensor{\varPhi};\tensor{x}) =  \IntBt U_t(\tensor{\varPhi},\tensor{f};\tensor{x}) \dv.
\label{eq:totalen}
\end{equation}

For the configurational problem, however, the stationarity of the total potential energy via the material variation at fixed spatial placement is not guaranteed, i.e. $\delta_x \pi \neq 0$. This is best understood by considering an example of a crack propagation problem, where a propagating crack is an irreversible change in the material configuration itself, resulting in the release of stored energy.

\subsection{Energy release and Eshelby stress}

To produce spontaneous (negative) energy release $\delta_x \pi \leq 0$, configurational tractions must oppose the material configurational changes. The material variation of Eq. \eqref{eq:totalen} produces
\begin{equation}
\begin{split}
\delta_x \pi =& \IntBt \partial_{\tensor{f}} U_t : \delta_x \tensor{f} \dv + \IntBt \partial_{\tensor{\varPhi}} U_t \cdot \delta \tensor{\varPhi} \dv \\
=& \IntBo \tensor{p} : \[\grad_X \delta \tensor{\varPhi} \cdot \tensor{f}\] J \dV - \IntBo \tensor{B}_t \cdot \[\delta \tensor{\varPhi}\] J \dV \\
=& \IntBo \tensor{\it\Sigma} : \grad_X \delta \tensor{\varPhi} \dV - \IntBo \tensor{B}_0 \cdot \delta \tensor{\varPhi} \dV \\
=& \IntSBo \delta \tensor{\varPhi} \cdot \tensor{\it\Sigma} \cdot \tensor{N} \dA - \IntBo \delta \tensor{\varPhi} \cdot \[ \text{Div} \tensor{\it\Sigma} + \tensor{B}_0 \] \dV,
\end{split}
\end{equation}
where $\tensor{p} := \partial_{\tensor{f}} U_t$ is a two-point (Piola-like) stress tensor with spatial reference and $\tensor{B}_t := -\partial_{\tensor{\varPhi}} U_t$ is the material body force with spatial reference. By Piola transformation we obtain a material stress tensor with material reference, the so-called Eshelby stress: $\tensor{\it\Sigma} = J\tensor{p} \cdot \tensor{f}^t$. For a detailed explanation of the relations between material and spatial references, see \cite{steinmann2000application, steinmann2009secret}. 

Due to the localized force balance $\text{Div} \tensor{\it\Sigma} + \tensor{B}_0 = \tensor{0}$, which can be related to $\text{Div} \tensor{P} + \tensor{b}_0 = \tensor{0}$ in Eq. \eqref{eq:localbalancep} (see \cite{steinmann2022spatial}), only the first term remains, representing the energy release due to configurational changes
\begin{equation}
\cR := \delta_x \pi = \IntSBo \delta \tensor{\varPhi} \cdot \tensor{T}_0 \dA \leq 0,
\label{eq:energyrelease}
\end{equation}
where the material traction is the projection of the Eshelby stress to the outwards-pointing normal
\begin{equation}
\tensor{T}_0 = \tensor{\it\Sigma} \cdot \tensor{N}.
\end{equation}

\begin{remark}\label{re:jint}
	An important observation that indicates the physical significance of the introduced energy release is its relation to the classical definition of the (vector-valued) $\tensor{\cJ}$-integral. This relation, as shown in \cite{steinmann2009secret} for a singularity inside $\cB_0$, is given by
	
	\begin{equation}
	\begin{split}
	\cR &= \lim_{r \rightarrow 0} \int_{\partial S_0}\  \delta_X \tensor{\varPhi} \cdot \tensor{\it\Sigma} \cdot \tensor{N} \ \mathrm{dA} \\
	&= [\lim_{r \rightarrow 0} \delta \tensor{\varPhi}] \cdot \[\lim_{r \rightarrow 0} \int_{\partial S_0}\ \tensor{\it\Sigma} \cdot \tensor{N} \ \mathrm{dA} \] \\
	&=: -[\lim_{r \rightarrow 0} \delta \tensor{\varPhi}] \cdot \tensor{\cJ}.
	\end{split}
	\end{equation}
	\label{eq:jint}
\end{remark}

The Eshelby stress can alternatively be represented in energy-momentum format \citep{eshelby1975elastic, hill1986energy} as follows
\begin{equation}
\tensor{\it\Sigma} = U_0 \tensor{I} - \tensor{F}^t \cdot \partial_{\tensor{F}} U_0.
\label{eq:enmomshelby}
\end{equation}
The beauty of this formulation is that the Eshelby stress can be computed directly from a spatial motion problem in a post-processing step. 

\subsection{Role of configurational forces}

The configurational problem, which leads to the concept of energy release (Eq. \eqref{eq:energyrelease}) and its close relation to the well-known $\tensor{\cJ}$-integral (Remark \ref{re:jint}), forms a basis for the so-called Configurational Force Method (\textit{alt:} Material Force Method). To compose discrete configurational forces, we consider a discrete (FE) system for the weak form of the configurational problem with material reference
\begin{equation}
\asse_e^{n_{\rm el}} \IntBoh \tensor{\it\Sigma} \cdot \grad_X \cN^i - \tensor{B}_0\ \cN^i \dV = \asse_e^{n_{\rm el}} \IntSBoh \cN^i\ \tensor{T}_0 \dA,
\end{equation}
where $\cN^i$ is an element shape function associated with the $i$th node and $\asse_e^{n_{\rm el}}$ is the assembly operator over all finite elements. Assuming a homogeneous material and no body forces, we arrive at the definition of a discrete configurational force for the $I$th global node
\begin{equation}
\tensor{\cF}^I_{\rm CNF} = \asse_e^{n_{\rm el}} \IntBoe \tensor{\it\Sigma} \cdot \grad_X \cN^i \dV.
\label{eq:cnf}
\end{equation}

Since configurational forces are the configurational equivalent of deformational forces in a deformational problem, they serve as the driving forces for the assessment and propagation of material defects, like cracks or dislocations. This is due to the fact that they are power conjugated to configurational changes of the material nodal positions $\delta \tensor{X} = \delta \tensor{\varPhi}\(\tensor{x}\)$. For a deeper understanding of the physical meaning of configurational forces, refer to \cite{maugin1995material, steinmann2009secret}. 

\begin{remark}\label{re:cnf}
In the context of fracture mechanics, where $\tensor{X}$ are understood as design variables (which enable configurational changes), configurational forces are equivalent to the sensitivities in linear elastic setting as described in \cite{van2010crack}.
\end{remark}

Configurational forces can be easily computed in the post-processing step of a deformational problem using the energy-momentum format of the Eshelby stress (Eq. \ref{eq:enmomshelby}). This is the basis of the Configurational Force Method. Some application examples of the Configurational Force Method include finite element mesh adaptivity \citep{mueller2004use, scherer2007energy}, fracture mechanics \citep{steinmann2000application, steinmann2001application, mueller2002material, nguyen2005material, moreno2024configurational}, or crystal defects \citep{steinmann2002spatial}.

\section{h-Adaptive meshing using configurational forces}\label{sec:method}
As shown in previous works \citep{thoutireddy2004variational, mueller2004use, scherer2007energy}, configurational forces appear to be an excellent criterion for mesh control within a finite element context. Firstly, they provide a physics-based information on potential configurational changes such as defect motion, being power conjugated to the material variations of a discretized structure. Secondly, as purely discrete (nodal) quantities, their values strictly depend on the finite element discretization, i.e. their magnitude is proportional to the size of elements contributing to a particular node. In general, we can say that configurational forces provide an information about potential energy release due to configurational changes within the finite element discretization.

As previously discussed, mesh adaptivity in the context of topology optimization has been addressed either for improved design resolution using density-based criteria or for computational accuracy and efficiency using error-based criteria. In the following, we exploit configurational forces for mesh adaptivity in topology optimization to simultaneously target improved design resolution and computational accuracy and efficiency. This is possible due to the nature of configurational forces, since all hypothetical defects (within homogeneous materials and without pre-existing defects) can only occur at the boundary (they indicate where the boundary is) and especially in highly stressed regions (they indicate where the computational accuracy is needed).

The strategy presented in this work addresses topology optimization using structured meshes. For problems involving unstructured meshes, in addition to h-adaptivity, r-adaptivity presents an attractive alternative, as it can leverage both the magnitude and orientation of configurational forces \citep{thoutireddy2004variational, scherer2007energy}.
\subsection{Eshelby stress in topology optimization}
The Eshelby stress (Eq. \eqref{eq:enmomshelby}) and consequently the configurational forces (Eq. \eqref{eq:cnf}) have to be adapted to the pseudo-density field of the SIMP (Solid Isotropic Material with Penalization) method \citep{bendsoe1989optimal, bendsoe1999material}. We use a variant of the $\varepsilon$ relaxation according to \cite{cheng1997varepsilon} and interpolate the Eshelby stress as follows
\begin{equation}
\tensor{\it\Sigma}_\rho(\hat\rho_e) = f_{\varepsilon}(\hat\rho) \tensor{\it\Sigma} = \frac{\hat\rho}{\varepsilon \[1 - \hat\rho\] + \hat\rho} \tensor{\it\Sigma},
\end{equation}
where $\hat\rho$ denotes a regularized pseudo-density $\rho$. The regularization procedure follows filtering through the Helmholtz-type PDE filter \citep{lazarov2011filters} with boundary energy term as proposed by \cite{wallin2020consistent} to account for improved behavior at the boundaries of the design domain. The smoothed Heaviside projection \citep{wang2011projection} is then applied to the filtered density $\tilde\rho$ to provide a crisp design representation. The Eshelby stress $\tensor{\it\Sigma}$ is calculated in a standard way according to Eq. \eqref{eq:enmomshelby}. The configurational forces then simply utilize the regularized Eshelby stress tensor
\begin{equation}
\tensor{\cF}^I_{\rm CNF} (\hat\rho) = \asse_e^{n_{\rm el}} \IntBoe \tensor{\it\Sigma}_\rho (\hat\rho_e) \cdot \grad_X \cN^i \dV.
\label{eq:cnfdens}
\end{equation}

\subsection{Refinement and coarsening criteria}

In the following we define simple criteria for the configurational force-based refinement and coarsening
\begin{equation}
\begin{split}
\text{Refinement:}\quad \lVert\tensor{\cF}_{\rm CNF}^I\rVert &\geq c_r\ \cF^{\rm max}_{{\rm CNF}}\\
\text{Coarsening:}\quad \lVert\tensor{\cF}_{\rm CNF}^I\rVert &\leq c_c\ \cF^{\rm max}_{{\rm CNF}},
\end{split}
\label{eq:ref}
\end{equation}
where $\cF^{\rm max}_{{\rm CNF}}$ is the maximum value of a configurational force in the current optimization iteration, i.e. it changes for every iteration. For each finite element $e$, the criterion is evaluated for the nodes $I$ belonging to it. The parameters $c_r$ and $c_c$ are user-defined multiplicators to set the threshold for refinement and coarsening, respectively. The adaptivity is implemented according to the following procedure:

\begin{algorithm}[h!]
	\SetAlgoVlined
	\KwIn{Number of elements $n_{el}$,\\ Number of vertices per element $n_{v}^e$}
	\KwOut{Refined and/or coarsened mesh}
	
	\For{$e \gets 1$ \KwTo $n_{el}$}{
		$cellRefine \gets \text{false}$\
		$cellCoarsenCounter \gets 0$\
		
		\For{$v \gets 1$ \KwTo $n_v^e$}{
			\uIf{$\lVert\tensor{\cF}_{\mathrm{CNF}}^v\rVert \geq c_r\ \cF^{\mathrm{max}}_{\mathrm{CNF}}$}{
				$cellRefine \gets \text{true}$\
			}
			\ElseIf{$\lVert\tensor{\cF}_{\mathrm{CNF}}^v\rVert \leq c_c\ \cF^{\mathrm{max}}_{\mathrm{CNF}}$}{
				$cellCoarsenCounter \gets cellCoarsenCounter + 1$\
			}
		}
		
		\uIf{$cellRefine$ \textbf{is} \texttt{true}}{
			Refine the cell\
		}
		\ElseIf{$cellCoarsenCounter = n_v^e$}{
			Coarsen the cell\
		}
	}
	\caption{Mesh adaptivity in topology optimization based on configurational forces.}
	\label{alg:ref}
\end{algorithm}

\begin{figure}[H]
	\centering
	\includegraphics[width=84mm]{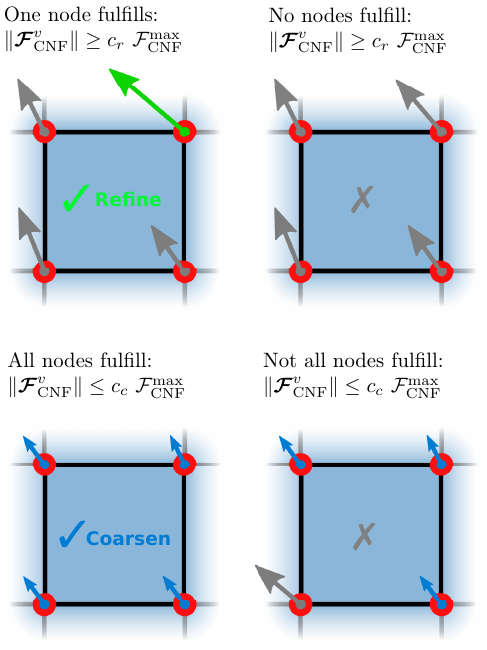}{}
	\caption{Possible cases of criteria fulfillment for refinement and coarsening based on Algorithm \ref{alg:ref}. A cell is marked for refinement if at least one of its nodes fulfills the refinement criterion from Eq. \eqref{eq:ref}, see the green force in the upper left image. A cell is marked for coarsening if all of its nodes satisfy the coarsening criterion from Eq. \eqref{eq:ref}, see the blue forces in the bottom-left image.}
	\label{fig:refinecase}
\end{figure}

In Algorithm \ref{alg:ref}, a finite element cell is marked for refinement if the magnitude of the configurational force at \textit{any} of its vertices satisfies the refinement criterion from Eq. \eqref{eq:ref}. However, a finite element cell is marked for coarsening only if the magnitude of the configurational force at \textit{all} of its vertices satisfies the coarsening criterion from Eq. \eqref{eq:ref}. In Fig. \ref{fig:refinecase} the logic of Algorithm \ref{alg:ref} is shown for better understanding.

\begin{figure}[H]
	\centering
	\includegraphics[width=65mm]{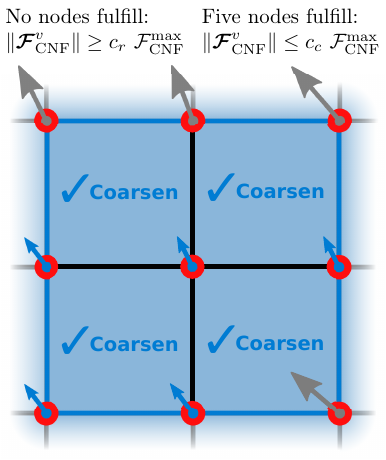}{}
	\caption{Coarsening is applied to four cells belonging to the same parent (inactive) cell, which is marked with a blue frame. If one of the child cells is marked for coarsening, deal.II takes care of marking the remaining three cells for coarsening, provided they were not marked for refinement.}
	\label{fig:coarsening}
\end{figure}

The reason why the coarsening criterion is more conservative (all nodes of a cell must satisfy the coarsening criterion instead of just one) is due to the simple fact that in a structured mesh, four adjacent cells belonging to a so-called parent cell must be coarsened at the same time, see Fig. \ref{fig:coarsening}. Thus, coarsening a cell affects a four times larger area and may affect regions where coarsening is not desired. Therefore, it seems reasonable to approach cell coarsening with more caution.

\begin{figure}[H]
	\centering
	\includegraphics[width=65mm]{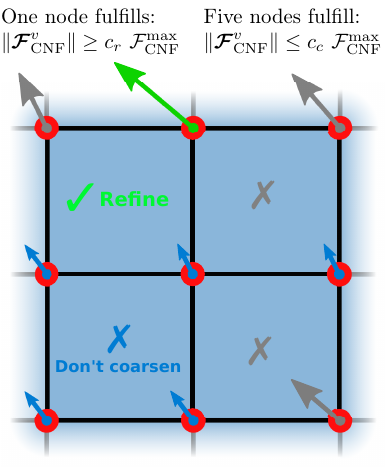}{}
	\caption{An edge case where two adjacent cells are marked for refinement and coarsening. In this case, only the refinement will occur, since the coarsening would affect the cell marked for refinement and therefore will not happen.}
	\label{fig:refinebeforecoarsen}
\end{figure}

According to the sequence of \textit{if} conditions in Algorithm \ref{alg:ref}, refinement has priority over coarsening. Fig. \ref{fig:refinebeforecoarsen} shows a case where this is relevant, i.e. where two cells belonging to the same parent cell (a potentially larger cell that would be "created" during coarsening) are marked for refinement and coarsening, respectively. This order follows the obvious reasoning that we prioritize geometric and numerical accuracy over computational efficiency.

\subsection{Handling multilevel coarsening and refinement}

Our structural optimization software utilizes the deal.II library \citep{bangerth2007deal} and its dependencies. Thus, the mesh model and compatibility aspects of the adaptive mesh are handled by deal.II internal functionalities. In general, the mesh object stores not only the active elements (actually used in the simulation), but also all its parent elements, representing the refinement history in a tree-like structure. This allows multi-level coarsening as the inactive parent elements are still available and can be reactivated, for more information see \cite{bangerth2007deal}.

\begin{figure}[t]
	\centering
	\includegraphics[width=65mm]{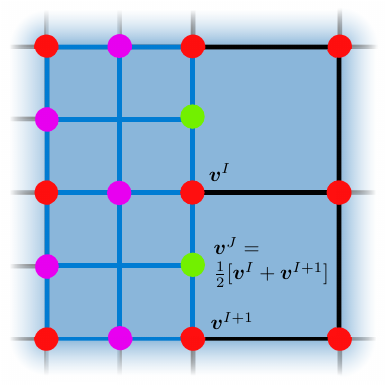}{}
	\caption{Adaptively refined mesh with hanging nodes (green). Red nodes: the unchanged, original nodes before refinement; purple nodes: newly created nodes with normal connectivity; green nodes: the hanging nodes that have no connection to the element of the lower refinement level.}
	\label{fig:hanging}
\end{figure}

Since the mesh is perfectly structured and consists entirely of quadrilateral elements, an adaptively refined and coarsened mesh must deal with hanging nodes (Fig. \ref{fig:hanging}). The hanging nodes are geometrically constrained by the neighboring nodes $\tensor{v}^J = \frac{1}{2}[\tensor{v}^I + \tensor{v}^{I+1}]$, where $\tensor{v}^J$ is the discrete solution at a newly introduced hanging node.

Compatibility criteria ensuring that refinement and coarsening do not lead to unwanted situations, such as adjacency between elements with a refinement level difference of two or the presence of so-called unrefined islands, are available and active by default. 

\section{Examples}\label{sec:examples}

To demonstrate the practical implications of configurational force-based (CNF-based) mesh adaptivity in topology optimization, two example studies are presented, a cantilever benchmark and a U-beam structure with stress singularity. To evaluate the merits of CNF-based mesh adaptivity, we compare it with density-based (DENS-based) and von Mises stress-based (VNM-based) adaptivity criteria. The following examples are modeled in nonlinear elastic setting, employing the Neo-Hookean material model with the Lam\'{e} constants: $\lambda = 2.66$ and $\mu = 0.71$. The finite element system applies biquadratic quadrilateral elements, resulting in 18 degrees of freedom per element. \footnote{A comparative study with linear elements (polynomial degree 1) revealed no significant differences in the optimized topologies or convergence behavior. As such, the detailed results for the linear element case are omitted here for brevity, since they do not offer additional insight into the performance of the proposed mesh adaptivity strategy beyond confirming its robustness with respect to the choice of element interpolation order.}

\subsection{Comparison criteria}

DENS-based mesh adaptivity is given by the following formula
\begin{equation}
\begin{split}
\text{Refinement:}\quad &0.2 \leq \tilde\rho_e \leq 0.8\\
\text{Coarsening:}\quad &\tilde\rho_e \leq 0.01\quad \text{or}\quad \tilde\rho_e \geq 0.99,
\end{split}
\end{equation}
where $\tilde\rho_e$ is the filtered density field in the element $e$. Naturally, DENS-based refinement is primarily concerned with the geometric accuracy of the design by targeting the grey transition regions within the design. By choosing the filtered density $\tilde\rho_e$, as opposed to the raw density $\rho$ or the projected density $\hat\rho_e$, we ensure the presence of grey densities ($0.2 \leq \tilde\rho_e \leq 0.8$) at the boundary of the design.

VNM-based mesh adaptivity criterion employs von Mises stress ($\sigma_{\rm VM}$) and targets numerical accuracy, especially for stress-constrained problems or structures with stress concentration regions. We choose the same strategy as for the CNF-based adaptivity of Eq. \eqref{eq:ref}
\begin{equation}
\begin{split}
\text{Refinement:}\quad \sigma_{\rm VM} &\geq c_r \sigma_{\rm VM}^{\rm max}\\
\text{Coarsening:}\quad \sigma_{\rm VM} &\leq c_c \sigma_{\rm VM}^{\rm max}.
\end{split}
\end{equation}
Here, the multiplicators $c_r$ and $c_c$ are the same as for CNF-based adaptivity. Our numerical experiments confirmed that reasonable values are $c_r = 0.25$ and $c_c = 0.01$ for both mesh adaptivity strategies.

\subsection{Cantilever}

\begin{figure}[tb!]
	\centering	
	\includegraphics[width=70mm]{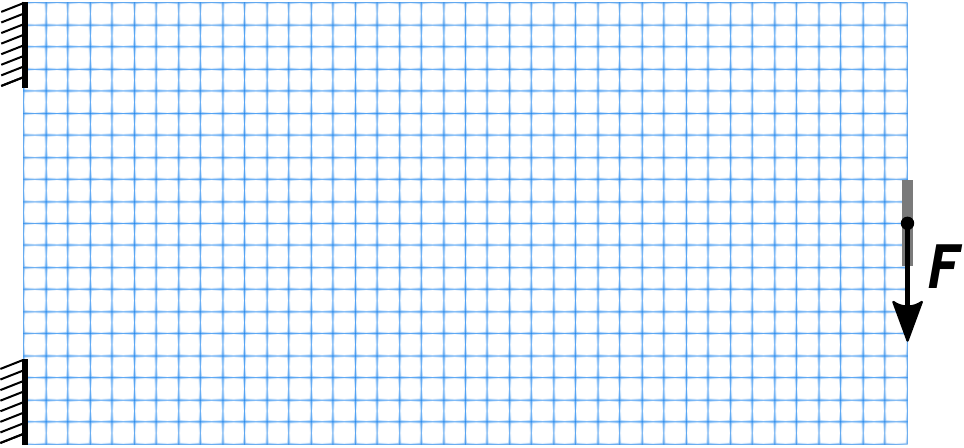}{}
	\caption{Setup of the cantilever benchmark and the initial mesh with a refinement level 1.} 
	\label{fig:cantileversetup}
\end{figure}

For CNF-based mesh adaptivity, it seems reasonable to target problems that contain stress concentrations and/or are stress constrained. This is due to the nature of configurational forces, which often serve as a criterion for configurational changes within a structure. However, to provide a better reference, we first consider the classic benchmark problem, which is the compliance minimization of a cantilever (see Fig. \ref{fig:cantileversetup}), given as
\begin{equation}
\begin{split}
\min_{\forall \rho}  \ &: \ \cF_{\textup{c}}\of{\rho, \tensor u} = \IntSBo \tensor u\of{\rho} \cdot \tensor{t}_0 \dA, \\
\textup{s.t.}\
&: \ \cG_{\textup{vol}}\of{\rho} = \IntBo \rho\of{\tensor{X}} \dV - \overline{V} \leq \ 0, \\
&: \ 0 \leq \ \rho_e \leq 1 \ \ \ e = 1,...,N_e,
\end{split}
\label{eq:compliance}
\end{equation}
where $\cF_{\textup{c}}\of{\rho, \tensor u}$ is the compliance functional, $\tensor u = \tensor{\varphi}\(\tensor X\) - \tensor X$ is the displacement and $\cG_{\textup{vol}}\of{\rho}$ is the volume function. To define the compliance, we assume the absence of body forces. For a discussion on well-posedness of Eq. \eqref{eq:compliance}, see \cite{bendsoe2013topology}.

\begin{figure}[tb!]
	\centering	
	\includegraphics[width=80mm]{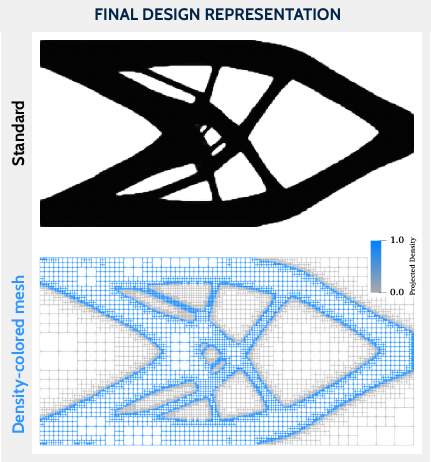}{}
	\caption{Example of a final design shown on top, using a black-to-white scale to fill the cells. In the bottom an alternative design representation features the mesh additionally. Here, a wireframe representation and a blue-to-gray color scale for the densities are chosen to maintain the visibility of the void regions and to avoid possible confusion with the black-to-white scale.} 
	\label{fig:designrepresentation}
\end{figure}

The following test cases are primarily concerned with the final mesh and the resulting geometric accuracy and computational efficiency. Therefore, the obtained designs are presented in the format shown in Fig. \ref{fig:designrepresentation}. That is, instead of showing the standard black and white density distribution, we have adopted a blue to grey scale for densities, which is applied to the element edges instead of the element fill. This way a clean representation of the final design is shown, displaying not only the final design (density distribution) but also the final mesh in both the solid \textit{and} void phase.

\begin{figure*}[tb!]
	\centering	
	\includegraphics[width=174mm]{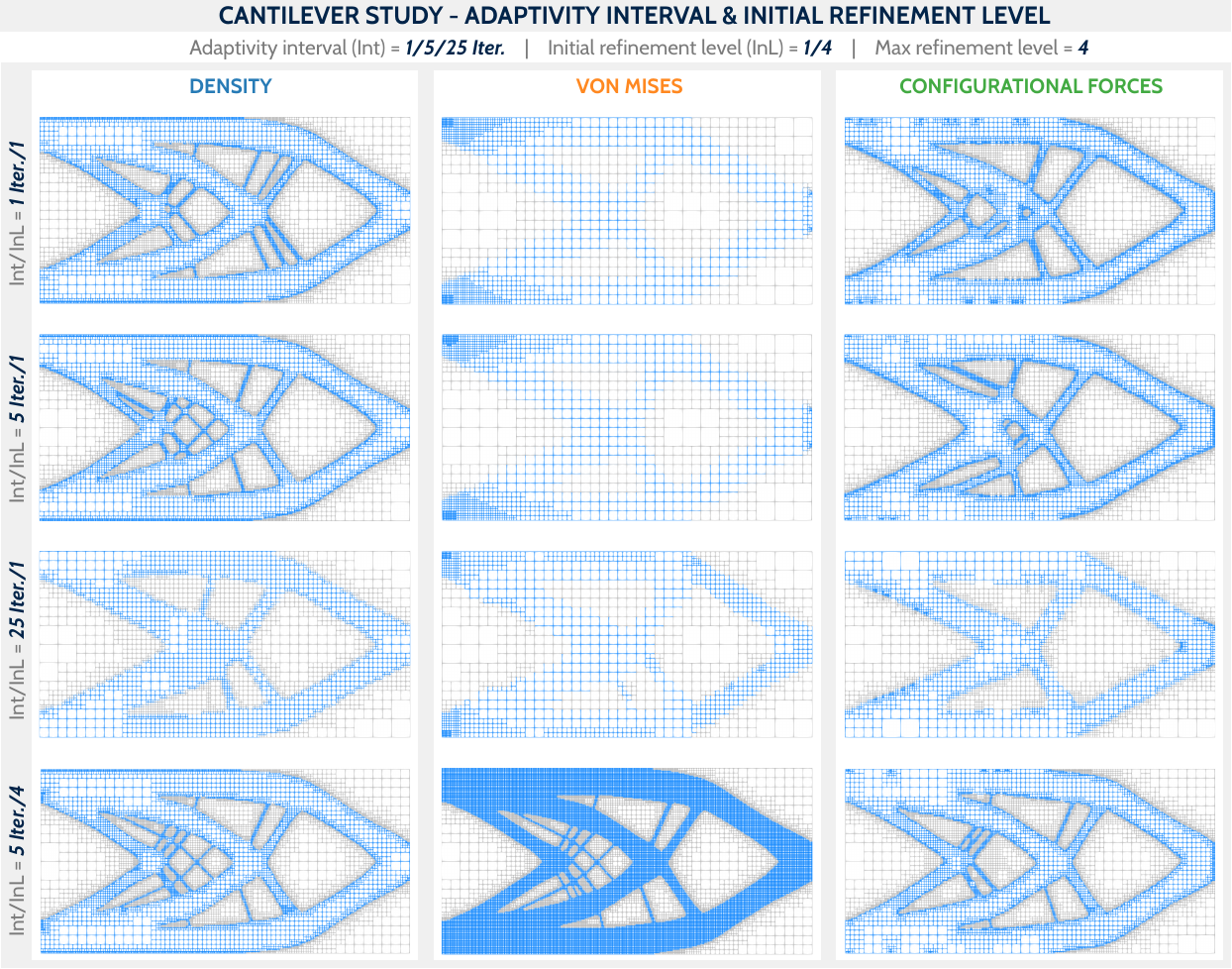}{}
	\caption{Final designs using the wireframe representation for the cantilever study. The columns are divided according to the mesh adaptivity technique and the rows according to the mesh adaptivity interval (Int) and the initial mesh refinement level (InL).} 
	\label{fig:cantileverstudy}
\end{figure*}

\begin{figure*}[tb!]
	\centering	
	\includegraphics[width=174mm]{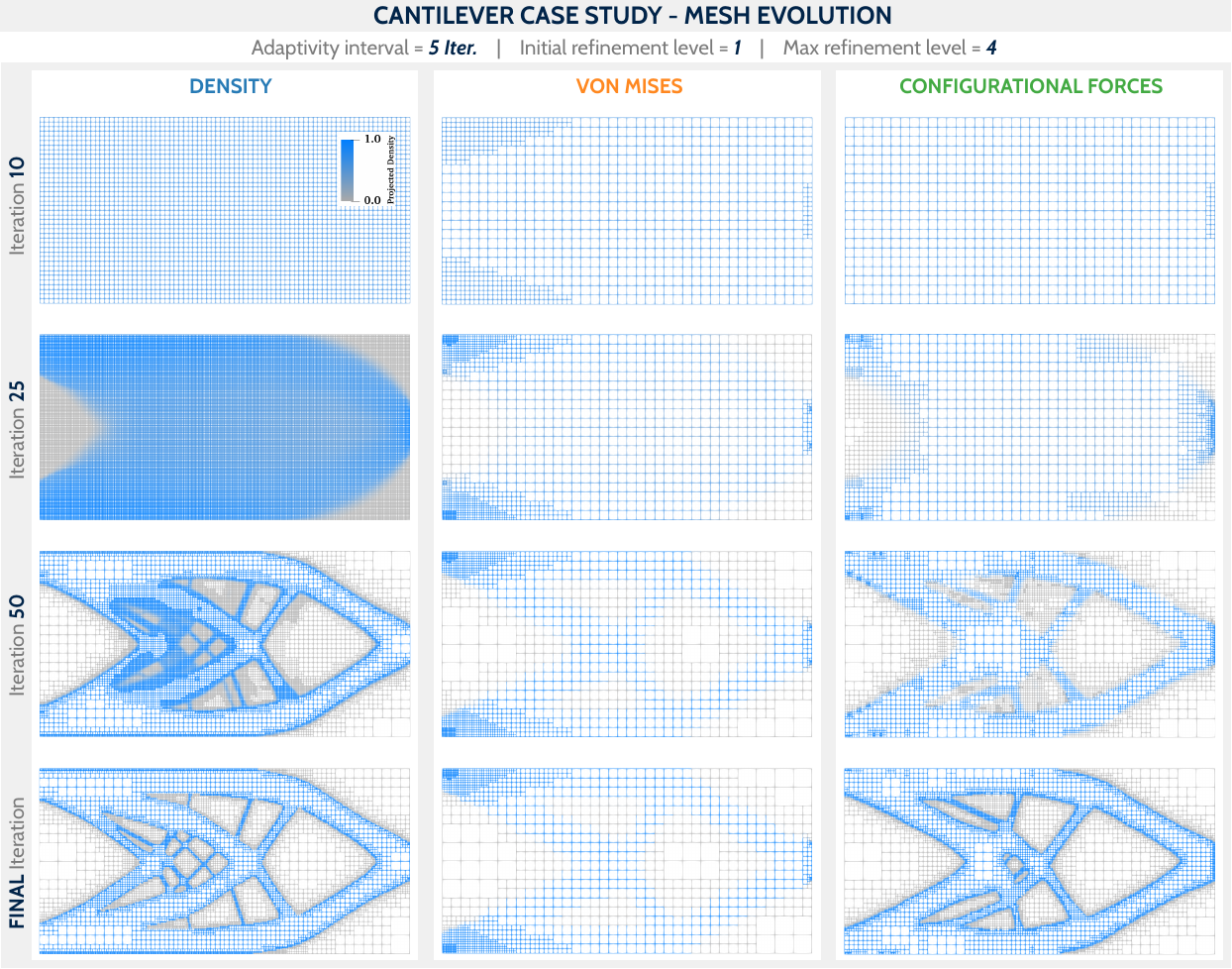}{}
	\caption{Mesh evolution for an adaptivity interval of 5 iterations. Due to the dominance of the volume constraint, the DENS-based criterion refines the mesh globally in the initial iterations, which consumes a significant amount of computational time.} 
	\label{fig:cantileverstudy1evo}
\end{figure*}

In the following study (see Table \ref{tab:cantilevercases}) we explored different intervals of mesh adaptation. That is, the mesh adaptation based on each of the above criteria takes place every $i$th iteration. 

\begin{table}[h]
	\centering
	\caption{Summary of the parameters studied in the cantilever benchmark. Three different cases of adaptivity interval and one case of fully refined initial mesh are compared. The row numbering corresponds to the images in Fig. \ref{fig:cantileverstudy}.}
	\label{tab:cantilevercases}
	\begin{tabular}{c|cc}
		\toprule
		\multicolumn{3}{c}{Common settings $r = 0.1$; $F = 0.001$; Max ref. level = 4} \\
		\midrule
		Case no. & Adaptivity interval & Initial ref. level\\ 
		\midrule
		Row 1 & 1 iteration & 1\\
		Row 2 & 5 iterations & 1\\
		Row 3 & 25 iterations & 1\\
		Row 4 & 5 iterations & 4\\
		\bottomrule
	\end{tabular}
\end{table}

For three different adaptivity intervals, i.e. every 1, 5, or 25 iterations, the initial refinement level is 1. The base mesh consists of 20x10 elements and does not allow further coarsening (it does not store parent cell information for level 0 elements). Using an initial refinement level of 1 renders a mesh of 40x20 elements and provides a good starting point for further refinement while still allowing for one level of coarser elements where needed. The maximum refinement level is limited to 4. In addition, a special case is studied where we start with a mesh at refinement level 4 with a mesh adaptation interval of 5 iterations. This case is particularly interesting because, to the best of our knowledge, such a multi-level coarsening problem has not been studied before in the context of topology optimization.

Fig. \ref{fig:cantileverstudy} shows the final results for the cases from Table \ref{tab:cantilevercases}. The DENS-based and CNF-based adaptivity strategies produce consistent results for all adaptivity intervals and initial refinement levels. The elements are refined along the boundaries of the structure and coarsened outside and inside the structure, which is desired for both geometric accuracy and computational efficiency. More frequent mesh refinement results in more detailed designs. The dependence of the geometric complexity on the adaptivity interval is pronounced for DENS-based and CNF-based adaptivity. The VNM-based mesh adaptivity strategy generally does not render geometrically refined structures because the stress values are highest near the Dirichlet and Neumann boundary conditions. In addition, it is highly dependent on the initial level of refinement, as for the fully refined initial structure it only coarsens void cells. However, it is important to keep in mind that VNM adaptivity is not a reasonable criterion for cantilever-type problems where the stress response is not relevant. The efficiency of VNM-based adaptivity is studied more carefully in the sequel for the U-beam problem.

\begin{figure}[bt!]
	\centering	
	\includegraphics[width=82mm]{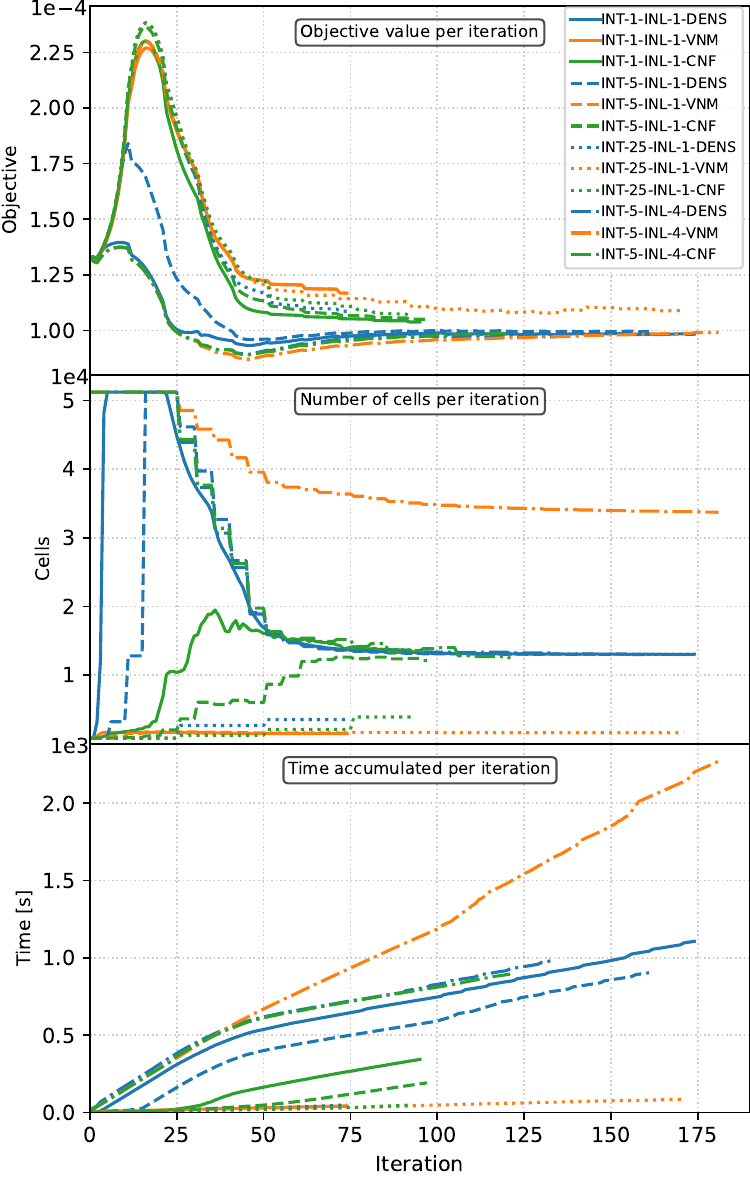}{}
	\caption{Evolution of the objective value (top), the number of cells (middle) and accumulated computation time (bottom) over optimization iterations for the cantilever benchmark. The coloring is assigned according to the mesh adaptivity method.} 
	\label{fig:cantileverplot}
\end{figure}

To address computational efficiency, the mesh evolution for selected iterations is shown in Fig. \ref{fig:cantileverstudy1evo} for the cases from the second row of Table \ref{tab:cantilevercases}, i.e. for the adaptivity interval of 5 iterations. In addition, the objective function (Fig. \ref{fig:cantileverplot}, top plot) the number of cells per iteration (Fig. \ref{fig:cantileverplot}, middle plot) and the accumulated computation time per iteration (Fig. \ref{fig:cantileverplot}, bottom plot) are plotted for all cases from Table \ref{tab:cantilevercases}. The analysis of the mesh evolution reveals the difference between DENS-based and CNF-based adaptivity. Due to the dominance of intermediate densities in the early iterations, the DENS-based adaptivity refines the entire mesh, which costs a significant amount of computation, as can be seen in Fig. \ref{fig:cantileverplot} (bottom plot) for the first 50 iterations. In the case of CNF-based adaptivity, few cells were initially refined, which did not significantly affect the performance. In the end, the CNF-based adaptivity case required only $21\%$ of the computational time of the DENS-based adaptivity case. The evolution and final values of the objective function show no major differences between the selected methodologies, with the only stand out case being the VNM-based adaptivity producing higher compliance values due to less refined boundaries.

\begin{figure}[bt!]
	\centering	
	\includegraphics[width=84mm]{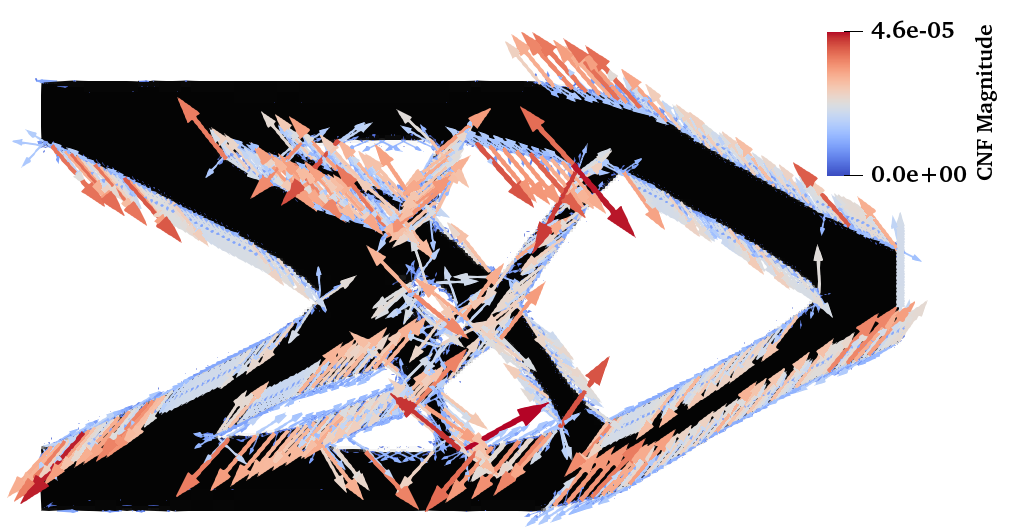}{}
	\caption{Configurational forces for the final cantilever design using CNF mesh adaptivity.} 
	\label{fig:cantileverforces}
\end{figure}

As a result of CNF mesh adaptivity, we obtain a uniform distribution of configurational forces along the design boundaries, as shown in Fig. \ref{fig:cantileverforces}. This is a particularly desirable state when fracture resistance is of interest.

A general conclusion from the cantilever study is that CNF-based mesh adaptivity is able to provide a well-defined geometry, similar to DENS-based mesh adaptivity, with improved robustness in the early iterations and resulting significant computational time saving. In contrast, VNM mesh adaptivity is not a reasonable choice for non-stress-constrained problems, so its comparison is more useful in the following example.

\subsection{U-beam}

\begin{figure}[tb!]
	\centering	
	\includegraphics[width=84mm]{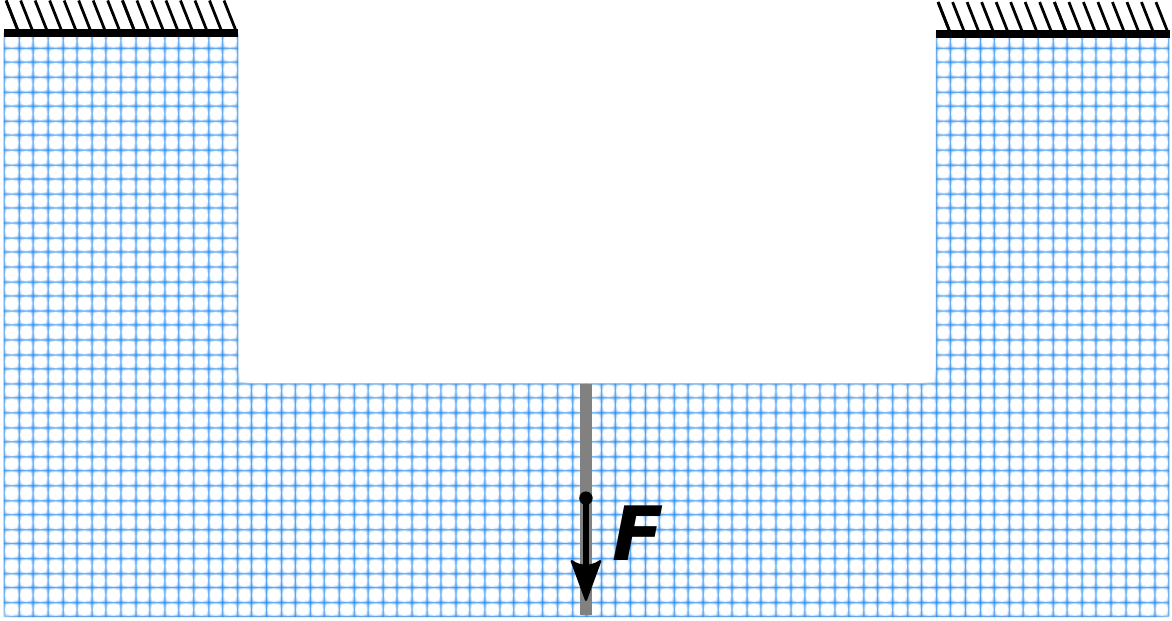}{}
	\caption{Setup of a U-beam example with an initial mesh of refinement level 1. By taking advantage of symmetry, only half of the structure is considered for optimization.} 
	\label{fig:ubeamsetup}
\end{figure}

A more sophisticated example in the context of configurational mechanics is a U-beam (Fig. \ref{fig:ubeamsetup}), which contains stress concentration regions (inner corners) that are natural candidates for material failure.

\begin{figure*}[tb!]
	\centering	
	\includegraphics[width=145mm]{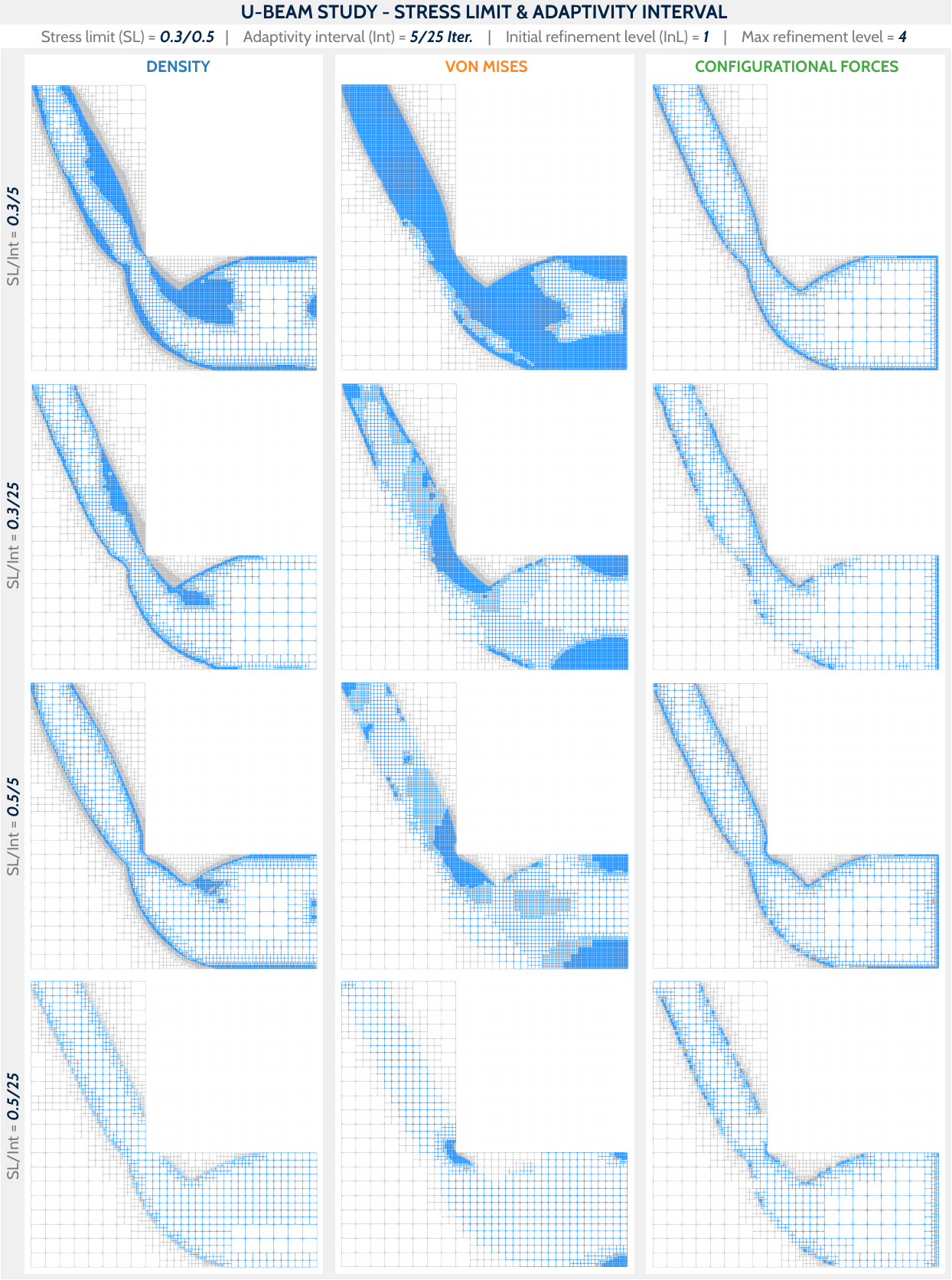}{}
	\caption{Final designs using the wireframe representation for the U-beam study. The columns are divided according to the mesh adaptivity technique and the rows according to the stress limit (SL) and mesh adaptivity interval (Int).} 
	\label{fig:ubeamstudy}
\end{figure*}

In this case, the introduction of a stress constraint in the optimization setup is essential to obtain a desirable structure that mitigates the stress buildup. Therefore, the optimization problem is defined as follows
\begin{equation}
\begin{split}
\min_{\forall \rho}  \ &: \ \cF_{\textup{c}}\of{\rho, \tensor u} = \IntSBo \tensor u\of{\rho} \cdot \tensor{t}_0 \dA, \\
\textup{s.t.}\
&: \ \cG_{\textup{vol}}\of{\rho} = \IntBo \rho\of{\tensor{X}} \dV - \overline{V} \leq \ 0, \\
&: \ \cG_{\textup{PVM}}\of{\rho, \tensor{u}} = \frac{1}{V^{\frac{1}{p}}} \left\Vert \sigma_N^p \right\Vert_p \leq \ 1, \\
&: \ 0 \leq \ \rho_e \leq 1 \ \ \ e = 1,...,N_e,
\end{split}
\end{equation}
where $\cG_{\textup{PVM}}$ is a volume normalized P-norm stress formulation, where the P-norm is given by
\begin{equation}
\cG_{\textup{PVM}}\of{\rho, \tensor{u}} = \left[ \frac{1}{\IntBo \dV} \ \IntBo \sigma_N^p \dV \right]^{\frac{1}{p}}
\end{equation}
and
\begin{equation}
\sigma_N^p = f_\varepsilon (\rho) \frac{\sigma_{VM}}{\sigma_{a}}
\end{equation}
is a stress measure that employs the $\varepsilon$-relaxed ($f_\varepsilon (\rho)$, \cite{cheng1997varepsilon}) von Mises stress $\sigma_{VM}$ normalized with respect to the user-defined stress limit $\sigma_{a}$, which is also referred to in this example by the abbreviation "SL". This formulation ensures that for a given stress limit $\sigma_{a}$, the stress constraint is satisfied if $\cG_{\textup{PVM}} \leq 1$.

This time, the case study focuses on the stress limit and the adaptivity interval as the main parameters, see Table \ref{tab:ubeamcases}. The initial refinement level of 1 is used for all cases.

\begin{table}[h!]
	\centering
	\caption{Summary of the parameters studied in the U-beam example. The row numbering corresponds to the images in Fig. \ref{fig:ubeamstudy}.}
	\label{tab:ubeamcases}
	\begin{tabular}{c|cc}
		\toprule
		\multicolumn{3}{c}{Common set.: $r = 0.2$; $F = 0.02$; Init/Max ref. level = 1/4} \\
		\midrule
		Case no. & Stress limit & Adaptivity interval\\ 
		\midrule
		Row 1 & 0.3 & 5 \\
		Row 2 & 0.3 & 25 \\
		Row 3 & 0.5 & 5 \\
		Row 4 & 0.5 & 25 \\
		\bottomrule
	\end{tabular}
\end{table}

\begin{figure}[bt!]
	\centering	
	\includegraphics[width=82mm]{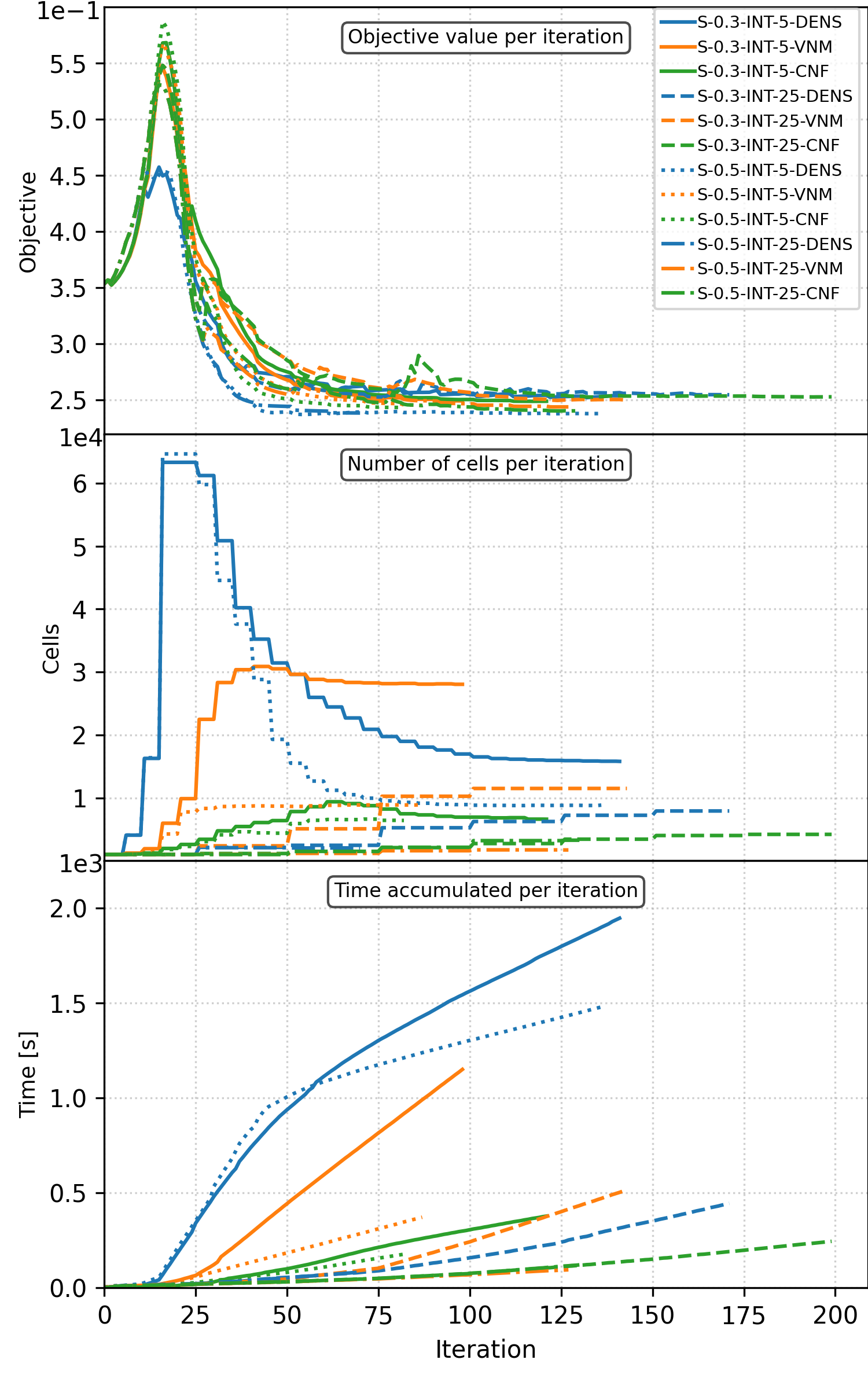}{}
	\caption{Evolution of the objective value (top), the number of cells (middle) and accumulated computation time (bottom) over optimization iterations for the U-beam. The coloring is assigned according to the mesh adaptivity method.} 
	\label{fig:ubeamplot}
\end{figure}

The final designs are shown in Fig. \ref{fig:ubeamstudy} and the corresponding plots for the objective function, the number of cells and the accumulated computation time per iteration are shown in Fig. \ref{fig:ubeamplot}, respectively. Upon visual inspection of the final designs in Fig. \ref{fig:ubeamstudy}, several observations can be made. DENS-based adaptivity is capable of generating geometrically detailed and smooth designs. However, for the top three cases, there were some artifacts within the solid phase, resulting in locally high refinement. The DENS-based adaptivity case with SL/Int = 0.5/25 (SL - stress limit) failed to refine the mesh around the stress concentration region, resulting in an inaccurate stress field and consequently failed to enforce a smooth geometry without a sharp corner. The VNM adaptivity naturally prioritized the highly stressed corner for refinement, but it did not render geometrically accurate boundaries along low stress regions. The stricter constraint value of SL = 0.3 results in larger fully refined regions, which affects the computation time. CNF-based adaptivity renders a consistent set of results, with full refinement only along the formed structure edges for the cases with Int = 5, guaranteeing both geometric and stress accuracy. Moreover, for the cases SL/Int = 0.5/25, DENS-based and VNM-based adaptivity produces complementary designs whose refinements add up to the design with CNF-based adaptivity. By evaluating the plots in Fig. \ref{fig:ubeamplot}, a similar tendency as for the cantilever benchmark is observed, i.e. a large computational cost for DENS adaptivity with the adaptivity interval Int = 5. The objective value plot (Fig. \ref{fig:ubeamplot}, top plot) depicts no significant differences for the final iterations between the studies cases. As expected, the less restrictive stress constraints of SL = 0.5 allowed for a slightly more improved value of the final compliance.

\begin{figure*}[tb!]
	\centering	
	\includegraphics[width=174mm]{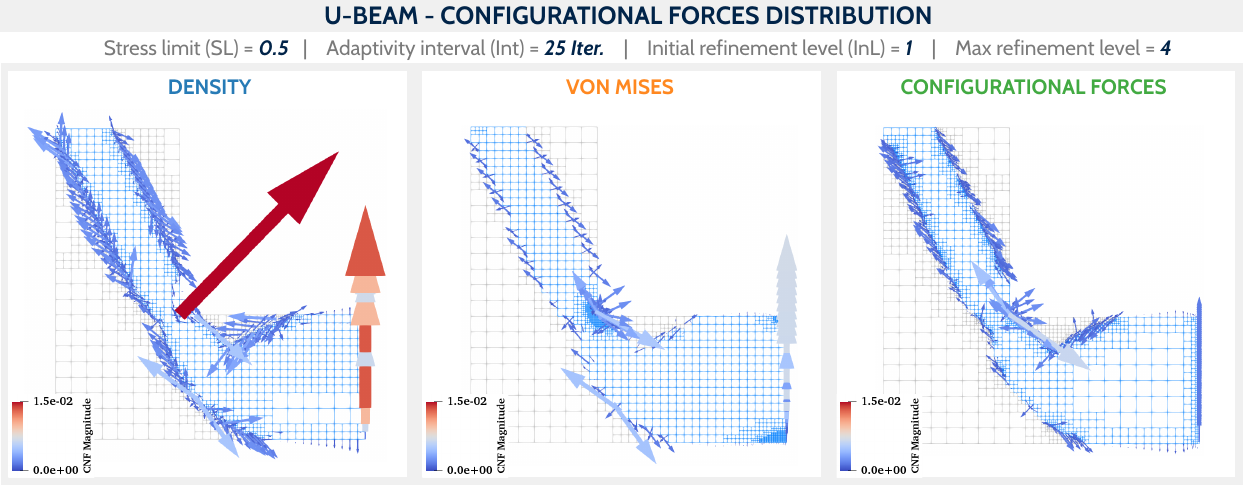}{}
	\caption{Configurational forces in the final designs for the setup SL/Int = 0.5/25, between the studied mesh adaptivity methods. For better comparison, the forces use the same scale in all cases.} 
	\label{fig:ubeamforces}
\end{figure*}

Due to the presence of stress concentration region, the U-beam is particularly susceptible to failure. Therefore, it is essential to analyze the configurational forces in the final designs. Fig. \ref{fig:ubeamforces} shows the final designs for the case s/Int = 0.5/25 together with the configurational forces. Their magnitudes are scaled consistently for all three examples. The dependence of the configurational forces on the element size is immediately apparent, especially for the DENS-based adaptivity. This highlights the importance of an equivalently refined mesh along the structure boundaries to allow a reasonable interpretation of the configurational forces. In the DENS-based adaptivity case, the largest magnitude of the force is naturally found around the unrefined stress concentration region. In the CNF-based adaptivity case, due to the consistently refined mesh along the stress concentration region, the distribution of forces is less variable in magnitude, but still peaks in magnitude in this corner.

\section{Conclusions and outlook}\label{sec:conclusions}

In order to obtain geometrically accurate designs with uncompromising computational cost, a novel combined mesh refinement and coarsening strategy based on configurational forces is introduced in the context of topology optimization. Configurational forces provide information about possible energy release due to configurational changes in a discrete system. Therefore, they identify critical regions in a structure that may be subject to failure due to fracture or dislocation motion. These regions usually require the highest accuracy and, consequently, an appropriate discretization. We have shown that, for topology optimization, mesh adaptivity based on configurational forces produces a highly refined mesh along the edges of the structure, providing targeted refinement for maximum geometric accuracy, similar to density-based mesh adaptivity. In addition, unlike density-based mesh refinement, high stress corners, in the example of the U-beam, are always maximally refined, ensuring highly accurate stress calculations. On the other hand, mesh refinement based on the von Mises stress value does not guarantee geometric accuracy along the edges of the structure as it only focuses on high stress regions. Furthermore, mesh adaptivity based on configurational forces is robust in terms of computational effort, providing consistent computational times regardless of the mesh adaptivity interval. Therefore, we can conclude that mesh adaptivity based on configurational forces combines the advantages of density-based and von Mises-based mesh adaptivity and is robust to different problem setups.

Since configurational forces are a vector field, they can be used for an anistropic type of mesh adaptivity, i.e., partial cell refinement along a selected direction, e.g., only along the x-axis. Future research will focus on anisotropic mesh adaptivity to further improve computational efficiency. In addition, mesh adaptivity based on configurational forces can be easily extended to three-dimensional problems. 
\bmhead{Acknowledgements}

The authors gratefully acknowledge financial support for this work by the European Research Council (ERC) under the Horizon Europe program (Grant-No.

101052785, project: SoftFrac).
\bmhead{Author contributions}
Conceptualization: Gabriel Stankiewicz; Methodology: Gabriel Stankiewicz; Implementation: Gabriel Stankiewicz and Chaitanya Dev; Discussions and improvement of the methodology: all authors; Writing - original draft preparation: Gabriel Stankiewicz; Writing - review and editing: all authors; Funding acquisition: Paul Steinmann, Resources: Paul Steinmann, Supervision: Paul Steinmann.
\bmhead{Funding}
This work is supported by the European Research Council (ERC) under the Horizon Europe program (Grant-No.
101052785, project: SoftFrac).
\bmhead{Data availability}
The version of the code, the executable, the parameter settings and the result files are available from the corresponding author
upon request.
\section*{Declarations}
\bmhead{Conflict of interest}
On behalf of all authors, the corresponding author
states that there is no conflict of interest.
\bmhead{Ethical approval}
This study does not involve human participants, animal subjects, or ethical concerns requiring institutional approval.
\bmhead{Replication of results}
All the presented methodology is implemented in C++ utilizing the finite element library deal.II \citep{bangerth2007deal, arndt2021deal}. The version of the code, the executable, the parameter settings and the result files are available from the corresponding author
upon request.

\begin{appendices}
\end{appendices}


\bibliography{reference-r1}

\end{document}